\documentclass{edm_article_modified}
\usepackage{etoolbox}
\usepackage{booktabs} 
\usepackage{graphicx}
\usepackage{wrapfig}
\usepackage{subcaption}
\usepackage{tikz}
\usetikzlibrary{arrows,positioning,shapes}
\tikzset{box/.style={rectangle, draw=black, minimum size=0.25cm}}
\usepackage{amsmath}

\usepackage{enumitem}
\usepackage{xcolor}
\usepackage{pifont}
\usepackage{colortbl}
\usepackage[numbers,sort&compress]{natbib}
\usepackage{hyperref}

\newcommand{\AlgoPEM}{\ensuremath{\textsc{PyFi-PEM}}}
\newcommand{\AlgoPyFiXNone}{\ensuremath{\textsc{PyFiX}_{\textnormal{shot:}\textsc{None}}}}
\newcommand{\AlgoPyFiXRand}{\ensuremath{\textsc{PyFiX}_{\textnormal{shot:}\textsc{Rand}}}}
\newcommand{\AlgoPyFiXSel}{\ensuremath{\textsc{PyFiX}_{\textnormal{shot:}\textsc{Sel}}}}
\newcommand{\AlgoPyFiXParallelSel}{\ensuremath{\textsc{PyFi||X}_{\textnormal{shot:}\textsc{Sel}}}}
\newcommand{\AlgoPyFiXVRule}{\ensuremath{\textsc{PyFiX-Rule}}}
\newcommand{\AlgoOurs}{\ensuremath{\textsc{PyFiXV}}}
\newcommand{\AlgoOursTitle}{P{\footnotesize \textbf{Y}}F{\footnotesize \textbf{I}}XV}
\newcommand{\AlgoPyFiXVOracle}{\ensuremath{\textsc{PyFiX-Opt}}}

\newcommand{\buggyprog}{\ensuremath{\mathcal{P}_{\textnormal{b}}}}
\newcommand{\fixedprog}{\ensuremath{\mathcal{P}_{\textnormal{f}}}}
\newcommand{\explaination}{\ensuremath{\mathcal{X}}}
\newcommand{\feedback}{\ensuremath{(\mathcal{P}_{\textnormal{f}}, \mathcal{X})}}
\newcommand{\dataFewS}{\ensuremath{\mathbb{D}_{\textnormal{shot}}}}
\newcommand{\dataVal}{\ensuremath{\mathbb{D}_{\textnormal{cal}}}}
\newcommand{\dataTest}{\ensuremath{\mathbb{D}_{\textnormal{test}}}}

\newcommand{\data}{\ensuremath{\mathbb{D}}}

\newcommand{\precision}{\ensuremath{\textnormal{P}}}

\definecolor{ExpHighlight}{rgb}{0.8, 0.1, 0.1}
\definecolor{mygreen}{rgb}{0,0.6,0}
\definecolor{mygray}{rgb}{0.5,0.5,0.5}
\definecolor{mymauve}{rgb}{0.58,0,0.82}
\definecolor{CodeHighlight}{rgb}{1,1,0.6}

\definecolor{PromtHeader}{rgb}{0.764,0.345,0.009}

\usepackage{listings}
\makeatletter
\let\old@lstKV@SwitchCases\lstKV@SwitchCases
\def\lstKV@SwitchCases#1#2#3{}
\makeatother
\usepackage{lstlinebgrd}
\makeatletter
\let\lstKV@SwitchCases\old@lstKV@SwitchCases

\lst@Key{numbers}{none}{%
    \def\lst@PlaceNumber{\lst@linebgrd}%
    \lstKV@SwitchCases{#1}%
    {none:\\%
     left:\def\lst@PlaceNumber{\llap{\normalfont
                \lst@numberstyle{\thelstnumber}\kern\lst@numbersep}\lst@linebgrd}\\%
     right:\def\lst@PlaceNumber{\rlap{\normalfont
                \kern\linewidth \kern\lst@numbersep
                \lst@numberstyle{\thelstnumber}}\lst@linebgrd}%
    }{\PackageError{Listings}{Numbers #1 unknown}\@ehc}}
\makeatother
\begin{document}

\title{Generating High-Precision Feedback for Programming Syntax Errors using Large Language Models\thanks{\textsuperscript{$1$}: Corresponding author.\\
\textsuperscript{\ \ $2$}: Listed in alphabetical order.\\
}
}


\numberofauthors{7}
\author{
\alignauthor
	 Tung Phung\textsuperscript{$1$}\\
      \affaddr{MPI-SWS}\\
      \email{mphung@mpi-sws.org}
\alignauthor
	 Jos{\'e} Cambronero\textsuperscript{$2$}\\
      \affaddr{Microsoft}\\
      \email{jcambronero@microsoft.com}
\alignauthor
	 \quad Sumit Gulwani\textsuperscript{$2$}\\
      \affaddr{\quad Microsoft}\\
      \email{\quad sumitg@microsoft.com}
\and
\alignauthor
	 Tobias Kohn\textsuperscript{$2$}\\
      \affaddr{TU Wien}\\
      \email{tobias.kohn@tuwien.ac.at	}
\alignauthor
	 Rupak Majumdar\textsuperscript{$2$}\\
      \affaddr{MPI-SWS}\\
      \email{rupak@mpi-sws.org}
\and
\alignauthor
	 Adish Singla\textsuperscript{$2$}\\
      \affaddr{MPI-SWS}\\
      \email{adishs@mpi-sws.org}
\alignauthor
	 Gustavo Soares\textsuperscript{$2$}\\
      \affaddr{Microsoft}\\
      \email{gsoares@microsoft.com}      
}

\maketitle


\begin{abstract}
Large language models (LLMs), such as Codex, hold great promise in enhancing programming education by automatically generating feedback for students. We investigate using LLMs to generate feedback for fixing syntax errors in Python programs, a key scenario in introductory programming. More concretely, given a student's buggy program, our goal is to generate feedback comprising a fixed program along with a natural language explanation describing the errors/fixes, inspired by how a human tutor would give feedback. While using LLMs is promising, the critical challenge is to ensure high precision in the generated feedback, which is imperative before deploying such technology in classrooms. The main research question we study is: \emph{Can we develop LLMs-based feedback generation techniques with a tunable precision parameter, giving educators quality control over the feedback that students receive?} To this end, we introduce \AlgoOurs, our technique to generate high-precision feedback powered by Codex. The key idea behind \AlgoOurs{} is to use a novel run-time validation mechanism to decide whether the generated feedback is suitable for sharing with the student; notably, this validation mechanism also provides a precision knob to educators. We perform an extensive evaluation using two real-world datasets of Python programs with syntax errors and show the efficacy of \AlgoOurs{} in generating high-precision feedback.
\end{abstract}


\vspace{-1mm}

\section{Introduction}
\label{sec.intro}

Large language models (LLMs) trained on text and code have the potential to power next-generation AI-driven educational technologies and drastically improve the landscape of computing education. One of such popular LLMs is OpenAI's Codex~\cite{DBLP:journals/corr/abs-2107-03374}, a variant of the $175$ billion parameter model GPT-3~\cite{DBLP:conf/nips/BrownMRSKDNSSAA20}, trained by fine-tuning GPT-3 on code from over $50$ million GitHub repositories. A recent study ranked Codex in the top quartile w.r.t. students in a large introductory programming course~\cite{DBLP:conf/ace/Finnie-AnsleyDB22}. Subsequently, recent works have shown promising results in using Codex on various programming education scenarios, including generating new programming assignments~\cite{DBLP:conf/icer/SarsaDH022}, providing code explanations~\cite{macneil23sigcse}, and enhancing programming-error-messages~\cite{leinonen23sigcse}.

We investigate the use of LLMs to generate feedback for programming syntax errors, a key scenario in introductory programming education. Even though such errors typically require small fixes and are easily explainable by human tutors, they can pose a major hurdle in learning for novice students~\cite{DBLP:conf/icer/PratherPMPHSC17}. Moreover, the programming-error-messages provided by the default programming environment are often cryptic and unable to provide explicable feedback to students~\cite{DBLP:conf/sigcse/Becker16,DBLP:conf/sigcse/KohnM20,DBLP:journals/cacm/Becker21}. Ideally, a human tutor would help a novice student by providing detailed feedback describing the errors and required fixes to the buggy program; however, it is extremely tedious/challenging to provide feedback at scale given the growing enrollments in introductory programming courses~\cite{singh2013automated,mirhosseini23sigcse}. To this end, our goal is to automate the feedback generation process using LLMs-based techniques. 

%

\begin{figure*}[t!]
\centering
	\begin{subfigure}[b]{.28\linewidth}
	\centering
	{
\begin{lstlisting}[basicstyle=\fontsize{6.5}{7.35}\ttfamily]
print "x"
a = input("xxx")
print  "Xx x"a
b = input("xxx xxx?")
print "xxxxxx" a + b
\end{lstlisting}
		\caption{Student's buggy program}				
		\label{fig.intro_overview.tj_pos.buggy}
	}
	\end{subfigure}
	\quad \ \ \ 
	\begin{subfigure}[b]{.28\linewidth}
	\centering
	{
\begin{lstlisting}[basicstyle=\fontsize{6.5}{7.35}\ttfamily,
        linebackgroundcolor={
            \ifnum\value{lstnumber}=3\color{CodeHighlight}\fi
            \ifnum\value{lstnumber}=5\color{CodeHighlight}\fi
        }]
print "x"
a = input("xxx")
print  "Xx x"+a
b = input("xxx xxx?")
print "xxxxxx" +a + b
\end{lstlisting}
		\caption{Generated fixed program}				
		\label{fig.intro_overview.tj_pos.fixed}
	}
	\end{subfigure}	
	\begin{subfigure}[b]{.255\linewidth}
	\centering
	{
         \scalebox{0.82}{
        \begin{tabular}{|p{1\linewidth}|}
     	  \hline
            \multicolumn{1}{|p{1\linewidth}|}{
                 The student forgets to concatenate the string literal with the variable. We can fix the error by adding a plus sign between the string literal and the variable.
            }\\
            \hline
        \end{tabular}
        }
        \vspace{2.2mm}
        \caption{Generated explanation}				
		\label{fig.intro_overview.tj_pos.exp}
	}
	\end{subfigure}
	\begin{subfigure}[b]{.11\linewidth}
	\centering
	{
		\includegraphics[height=1.60cm]{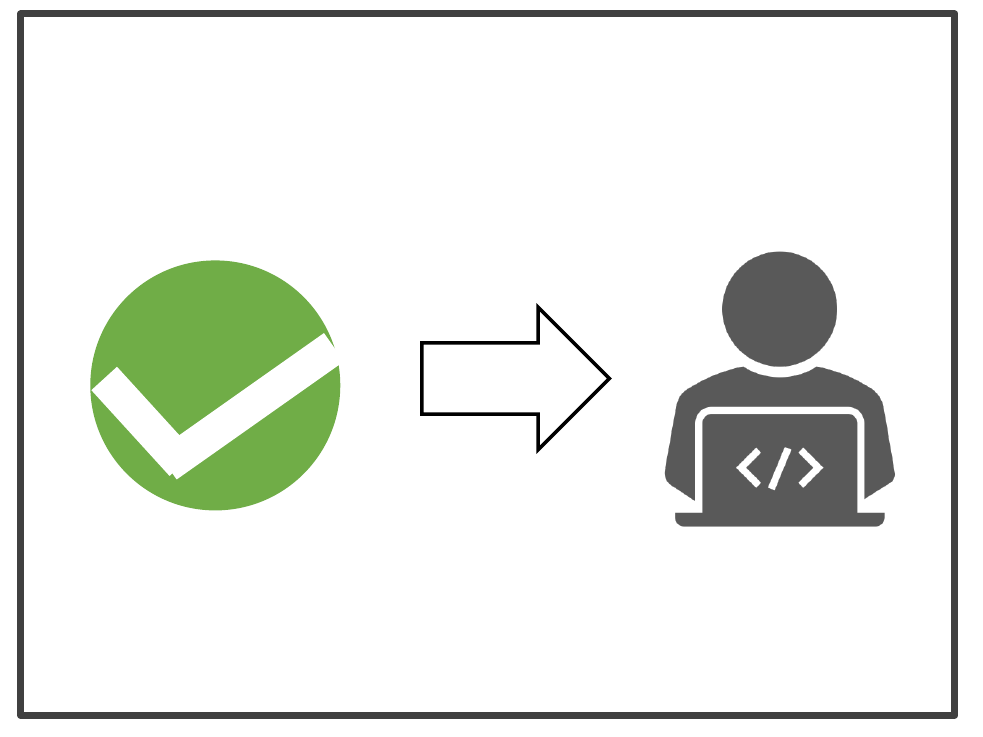}
        \vspace{-1.8mm}
		\caption{Validation}				
		 \label{fig.intro_overview.tj_pos.validation}
	}
	\end{subfigure}
    \caption{\looseness-1Illustrative example showcasing \AlgoOurs{} on a buggy Python $2$ program from TigerJython~\cite{DBLP:conf/sigcse/KohnM20}. \textbf{(a)} Buggy program (with string literals anonymized as sequences of `x' in our dataset).  \textbf{(b, c)} Fixed program and explanation generated by  \AlgoOurs{}. \textbf{(d)} \AlgoOurs{}'s validation mechanism \emph{successfully accepted} the generated feedback as suitable for sharing.
    }
	\label{fig.intro_overview.tj_pos}
\end{figure*}
%

\begin{figure*}[t!]
\centering
	\begin{subfigure}[b]{.28\linewidth}
	\centering
	{
\begin{lstlisting}[basicstyle=\fontsize{6.5}{6.7}\ttfamily]
for i in range(int(input()):
    l1=[]
    for i in range(int(input())):
        if i==0:
            if i==0:
                l1.appned(n)
            elif(i==(n-1)):
                l1.append(1)
            else:
                l1.append(i)
    print(l1)
\end{lstlisting}
		\caption{Student's buggy program}				
		\label{fig.intro_overview.cf_pos.buggy}
	}
	\end{subfigure}
	\quad \ \ \    
	\begin{subfigure}[b]{.28\linewidth}
	\centering
	{
\begin{lstlisting}[basicstyle=\fontsize{6.5}{6.7}\ttfamily,
        linebackgroundcolor={
            \ifnum\value{lstnumber}=1\color{CodeHighlight}\fi
            \ifnum\value{lstnumber}=6\color{CodeHighlight}\fi
        }]
for i in range(int(input())):
    l1=[]
    for i in range(int(input())):
        if i==0:
            if i==0:
                l1.append(n)
            elif(i==(n-1)):
                l1.append(1)
            else:
                l1.append(i)
    print(l1)
\end{lstlisting}
		\caption{Generated fixed program}				
		\label{fig.intro_overview.cf_pos.fixed}
	}
	\end{subfigure}
	\begin{subfigure}[b]{.255\linewidth}
	\centering
	{
        \scalebox{0.82}{
        \begin{tabular}{|p{1\linewidth}|}
     	  \hline
            \multicolumn{1}{|p{1\linewidth}|}{
                In line 1, a closing parenthesis is missing when calling function `range'. We should add a closing parenthesis at the end of line 1, before the colon, to fix this error. In line 6, the function `append' is misspelled. We should replace `appned' with `append' to fix this error.
                \vspace{1.2mm}
            }\\
            \hline
        \end{tabular}
        }
        \vspace{2mm}
        \caption{Generated explanation}				
		\label{fig.intro_overview.cf_pos.exp}
	}
	\end{subfigure}	
	\begin{subfigure}[b]{.11\linewidth}
	\centering
	{
		\includegraphics[height=2.94cm]{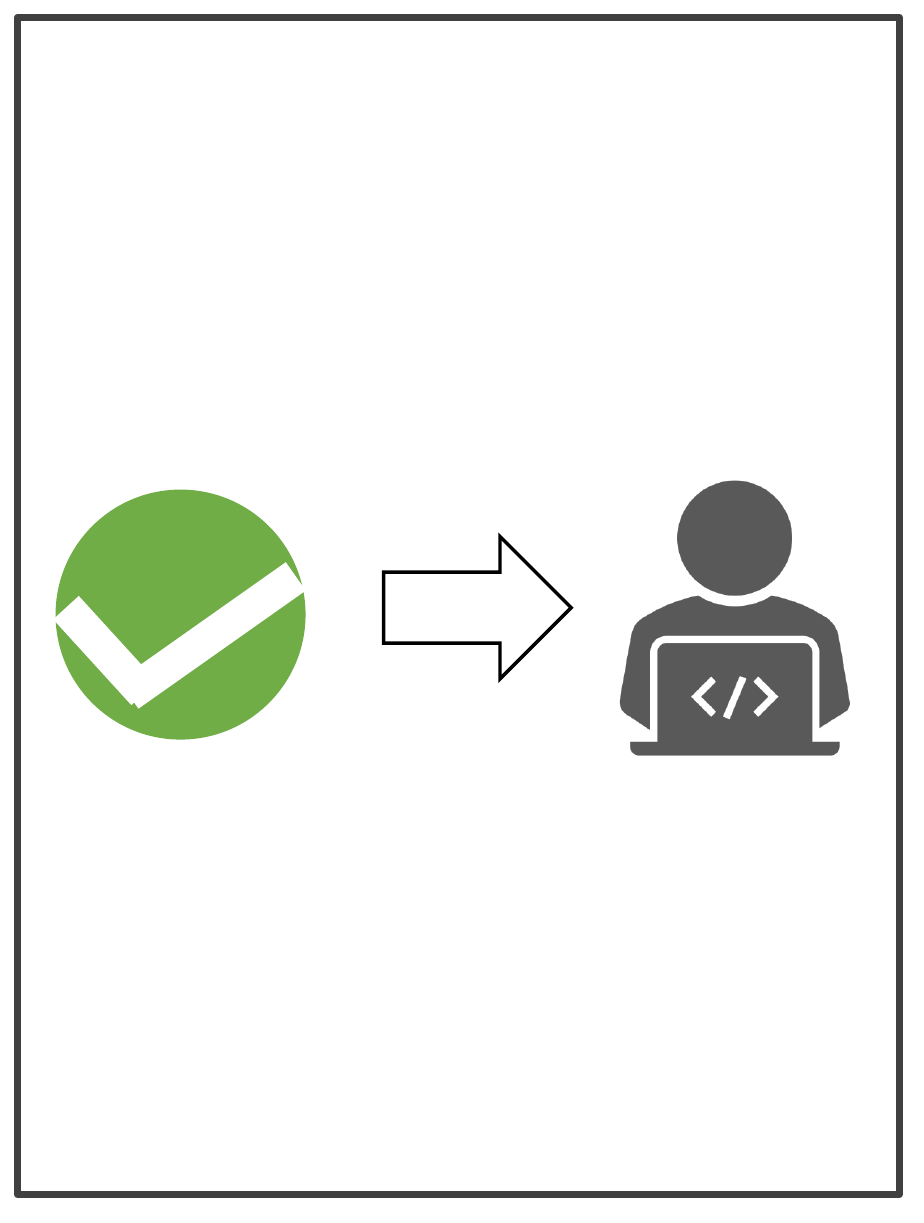}
        \vspace{-2.2mm}
		\caption{Validation}				
		 \label{fig.intro_overview.cf_pos.validation}
	}
	\end{subfigure}
    \caption{Similar to Figure~\ref{fig.intro_overview.tj_pos}, this example showcases \AlgoOurs{} on a buggy Python $3$ program from Codeforces~\cite{codeforces}.
    While the misspelling of `append' as `appned' is not a syntax error (raising an AttributeError at runtime instead), \AlgoOurs{} typically fixes such mistakes as well, showing the generalization value in LLMs-powered technique.
    }
	\vspace{-0.5mm}	
	\label{fig.intro_overview.cf_pos}
\end{figure*}

\begin{figure*}[t!]
\centering
	\begin{subfigure}[b]{.28\linewidth}
	\centering
	{
\begin{lstlisting}[basicstyle=\fontsize{6.5}{6.95}\ttfamily]
import math

def binary(num):
    x=int(math.log2(num))+1
    for i in range(x):
        num=(num ^^(1<<i))
        print(num)
for i in range(int,input()):
    n=int(input())
    binary(n)
\end{lstlisting}
		\caption{Student's buggy program}				
		\label{fig.intro_overview.cf_neg.buggy}
	}
	\end{subfigure}
	\quad \ \ \  
	\begin{subfigure}[b]{.28\linewidth}
	\centering
	{
\begin{lstlisting}[basicstyle=\fontsize{6.5}{6.95}\ttfamily,
        linebackgroundcolor={
            \ifnum\value{lstnumber}=6\color{CodeHighlight}\fi
            \ifnum\value{lstnumber}=8\color{CodeHighlight}\fi
        }]
import math

def binary(num):
    x=int(math.log2(num))+1
    for i in range(x):
        num=(num ^ (1<<i))
        print(num)
for i in range(int(input())):
    n=int(input())
    binary(n)
\end{lstlisting}
		\caption{Generated fixed program}				
		\label{fig.intro_overview.cf_neg.fixed}
	}
	\end{subfigure}
	\begin{subfigure}[b]{.255\linewidth}
	\centering
	{
        \scalebox{0.82}{
        \begin{tabular}{|p{1\linewidth}|}
     	  \hline
            \multicolumn{1}{|p{1\linewidth}|}{
                In line 6, the operator `\^{}\^{}' is not a valid operator in Python 3. We should replace it with `\^{}' to fix this error. \textcolor{ExpHighlight}{In line 8, a closing parenthesis is missing when calling function `range'. We should add a closing parenthesis at the end of line 8, before the colon, to fix this error.}
                \vspace{-4mm}
            }\\\\
            \hline
        \end{tabular}
        }
        \vspace{1.8mm}
        \caption{Generated explanation}				
		\label{fig.intro_overview.cf_neg.exp}
	}
	\end{subfigure}	
	\begin{subfigure}[b]{.11\linewidth}
	\centering
	{
		\includegraphics[height=2.80cm]{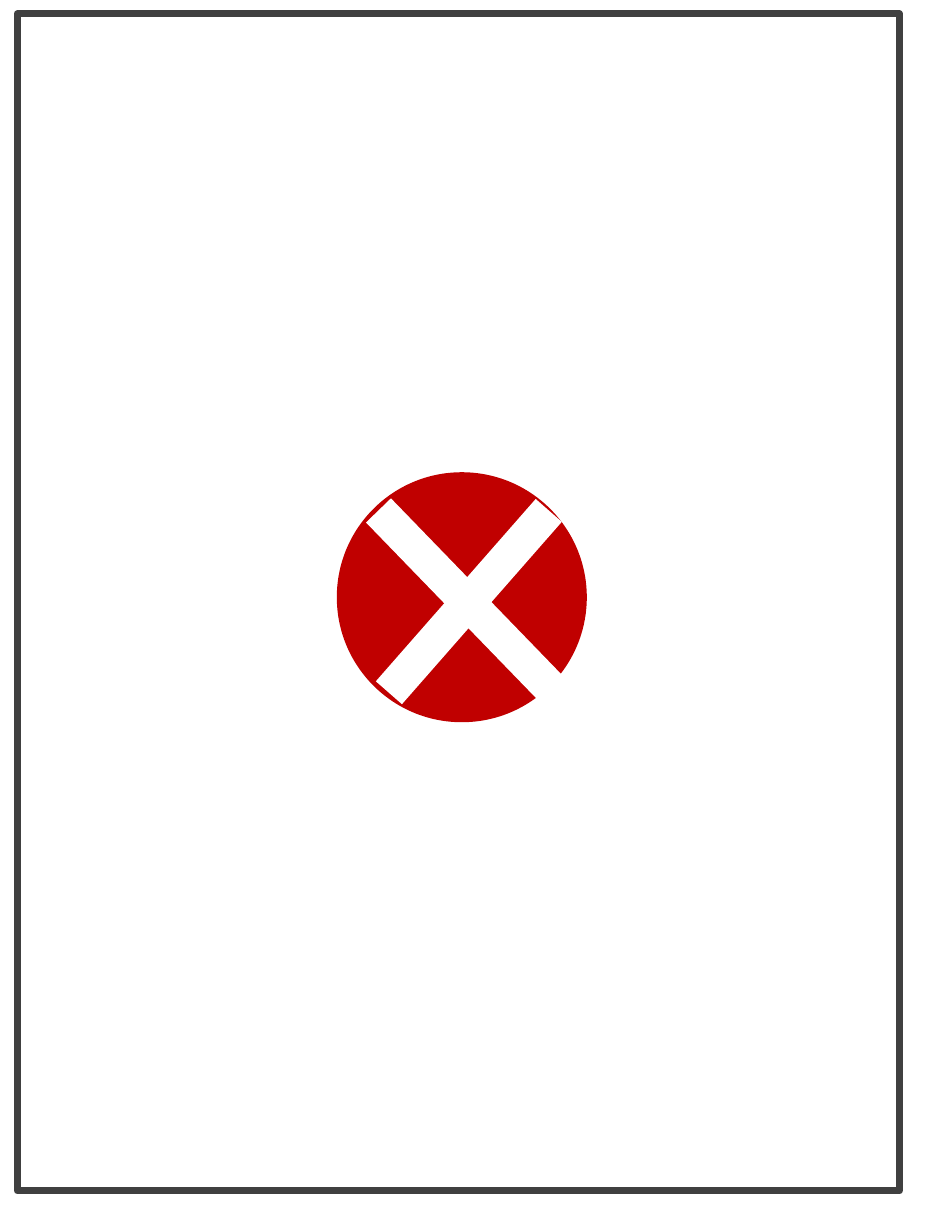}
        \vspace{-2.2mm}
		\caption{Validation}				
		\label{fig.intro_overview.cf_neg.validation}
	}
	\end{subfigure}
    \caption{Similar to Figure~\ref{fig.intro_overview.cf_pos}, this example showcases \AlgoOurs{} on a buggy Python $3$ program from Codeforces~\cite{codeforces}. \AlgoOurs{}'s validation mechanism \emph{successfully rejected} the generated feedback (we marked text in \textbf{(c)} to highlight issues with explanation).
    }
	\vspace{-1.5mm}
	\label{fig.intro_overview.cf_neg}
\end{figure*}

More concretely, given a student's buggy program, we want to generate feedback comprising a fixed program and a natural language explanation describing the errors/fixes, inspired by how a human tutor would give feedback. While models like Codex, trained on both text and code, are naturally suitable for this, the critical challenge is to ensure high precision in the generated feedback. High precision is imperative in building educators' trust before deploying such an AI-driven technology in classrooms. A recent work investigated enhancing the default programming-error-messages using Codex~\cite{leinonen23sigcse}; one of the takeaways, quoted from their paper, is \emph{``The key implications of this work are that programming error message explanations and suggested fixes generated by LLMs are not yet ready for production use in introductory programming classes...''}. Our initial experiments (Section~\ref{sec.experiments}) also highlight issues in generating high-precision feedback. To this end, the main research question is:

\vspace{-1.5mm}
\emph{Can we develop LLMs-based feedback generation techniques with a tunable precision parameter, giving educators quality control over the feedback that students receive?}

\vspace{-1.5mm}
\subsection{Our Approach and Contributions}
\label{sec.intro.contributions}

In this paper, we develop \AlgoOurs{}, our technique to generate high-precision feedback powered by Codex. Given a student's buggy program as input, \AlgoOurs{} decomposes the overall process into (i) feedback generation (i.e., a fixed program and a natural language explanation for errors/fixes); and (ii) feedback validation (i.e., deciding whether the generated feedback is suitable for sharing with the student). One of the key ideas in \AlgoOurs{} is to use a run-time feedback validation mechanism that decides whether the generated feedback is of good quality. 
This validation mechanism uses Codex as a \emph{simulated student model} -- the intuition is that a good quality explanation, when provided as Codex's prompt instruction, should increase Codex's success in converting the buggy program to the fixed program.
Notably, this validation also provides a tuneable precision knob to educators to control the precision and coverage trade-off. The illustrative examples in Figures~\ref{fig.intro_overview.tj_pos},~\ref{fig.intro_overview.cf_pos}, and \ref{fig.intro_overview.cf_neg} showcase \AlgoOurs{} on three different student's buggy programs. 
Our main contributions are:

\vspace{-2mm}
\begin{enumerate}[label=(\Roman*),parsep=1.5pt]
\item We formalize the problem of generating high-precision feedback for programming syntax errors using LLMs, where feedback comprises a fixed program and a natural language explanation. (Section~\ref{sec.problem})
\item We develop a novel technique, \AlgoOurs{}, that generates feedback using Codex and has a run-time feedback validation mechanism to decide whether the generated feedback is suitable for sharing.  (Section~\ref{sec.methods})
\item We perform extensive evaluations using two real-world datasets of Python programs with syntax errors and showcase the efficacy of \AlgoOurs{}. We publicly release the implementation of \AlgoOurs{}. (Section~\ref{sec.experiments})\footnote{Github: \url{https://github.com/machine-teaching-group/edm2023_PyFiXV} \label{footnote.githublink}}
%
\end{enumerate}
%

\subsection{Related Work}\label{sec.intro.related}
\textbf{Feedback generation for programming errors.} There has been extensive work on feedback generation for syntactic/semantic programming errors~\cite{DBLP:conf/pldi/GulwaniRZ18,DBLP:conf/icse/BhatiaKS18,DBLP:conf/aaai/GuptaKS19,DBLP:journals/corr/abs-2209-14876,joshi23aaai}; however, these works have  focused on fixing/repairing buggy programs without providing explanations. The work in \cite{singh2013automated} proposed a technique to generate explanations; however, it requires pre-specified rules that map errors to explanations. Another line of work, complementary to ours, has explored crowdsourcing approaches to obtain explanations provided by other students/tutors~\cite{DBLP:conf/chi/HartmannMBK10,DBLP:conf/lats/HeadGSSFDH17}. There has also been extensive work on improving the programming-error-messages by designing customized  environments~\cite{DBLP:conf/sigcse/KohnM20,DBLP:journals/cacm/Becker21}. As discussed earlier, a recent study used Codex to enhance these error messages~\cite {leinonen23sigcse}; however, our work is different as we focus on generating high-precision feedback with a tuneable precision knob.
%

\textbf{Validation of generated content.} In recent work, \cite{DBLP:journals/corr/abs-2210-00848} developed a technique to validate LLMs' output in the context of program synthesis. While similar in spirit, their validation mechanism is different and operates by asking LLMs to generate predicates for testing the synthesized programs. Another possible approach is to use back-translation models to validate the generated content~\cite{DBLP:conf/emnlp/EdunovOAG18,DBLP:conf/nips/PuEK0S20}; however, such a back-translation model (that generates buggy programs from explanations) is not readily available for our setting. Another approach, complementary to ours, is to use human-in-the-loop for validating low confidence outputs~\cite{DBLP:conf/aied/FunayamaSMMSI22}. 

%

\vspace{-0.2mm}
\section{Problem Setup}\label{sec.problem}
Next, we introduce definitions and formalize our objective.

\vspace{-1.3mm}
\subsection{Preliminaries}\label{sec.problem.preliminaries}
\textbf{Student's buggy program.} Consider a student working on a programming assignment who has written a buggy program with syntax errors, such as shown in 
Figures~\ref{fig.intro_overview.tj_pos.buggy},~\ref{fig.intro_overview.cf_pos.buggy}, and \ref{fig.intro_overview.cf_neg.buggy}. Formally, these  syntax errors are defined by the underlying parser of the programming language~\cite{DBLP:conf/pldi/GulwaniRZ18}; we will use the Python programming language in our evaluation. Henceforth, we denote such a buggy program as \buggyprog{}, which is provided as an input to feedback generation techniques.

\textbf{Feedback style.} Given \buggyprog{}, we seek to generate feedback comprising a fixed program along with a natural language explanation describing the errors and fixes. This feedback style is inspired by how a human tutor would give feedback to novice students in introductory programming education~\cite{DBLP:conf/sigcse/KohnM20,macneil23sigcse}. We denote a generated fixed program as \fixedprog{}, a generated explanation as \explaination{}, and generated feedback as a tuple \feedback{}.

\textbf{Feedback quality.} We assess the quality of generated feedback \feedback{} w.r.t. \buggyprog{} along the following binary attributes: (i) \fixedprog{} is syntactically correct and is obtained by making a small number of edits to fix \buggyprog{}; (ii) \explaination{} is complete, i.e., contains information about all errors and required fixes; (iii)  \explaination{} is correct, i.e., the provided information correctly explains errors and required fixes; (iv) \explaination{} is comprehensible, i.e., easy to understand, presented in a readable format, and doesn't contain redundant information. These attributes are inspired by evaluation rubrics used in literature~\cite{DBLP:conf/edm/ZhiMDLPB19,DBLP:conf/aied/GhoshTDS22,tack23edm,leinonen23sigcse}. In our evaluation, feedback quality is evaluated via ratings by experts along these four attributes. We measure feedback quality as binary by assigning the value of $1$ (good quality) if it satisfies \emph{all} the four quality attributes and otherwise $0$ (bad quality).\footnote{\looseness-1We note that the four attributes are independent. In particular, the attribute ``complete'' captures whether the explanation contains information about all errors/fixes (even though the information could be wrong), and the attribute ``correct'' captures the correctness of the provided information.
}

%

\subsection{Performance Metrics and Objective}\label{sec.problem.objectives}

\textbf{Performance metrics.} Next, we describe the overall performance metrics used to evaluate a feedback generation technique. For a buggy program \buggyprog{} as input, we seek to design techniques that generate feedback \feedback{} and also decide whether the generated feedback is suitable for sharing with the student. We measure the performance of a technique using two metrics: (i) \emph{Coverage} measuring the percentage number of times the feedback is \emph{generated and provided to the student}; (ii) \emph{Precision} measuring the percentage number of times the \emph{provided feedback is of good quality} w.r.t. the binary feedback quality criterion introduced above. In our experiments, we will compute these metrics on a dataset $\dataTest{} = \{\buggyprog{}\}$ comprising a set of students' buggy programs.\footnote{When a technique cannot generate feedback for an input program \buggyprog{} (e.g., the technique is unable to find a fixed program), then we use a natural convention that no feedback is provided to the student---this convention lowers the coverage metric but doesn't directly affect the precision metric.\label{footnote.nofeedback}}

\textbf{Objective.} Our goal is to design feedback generation techniques with high precision, which is imperative before deploying such techniques in classrooms. In particular, we want to develop techniques with a tuneable precision parameter that could provide a knob to educators to control the precision and coverage trade-off.
%

%

\begin{figure}[t!]
\centering
	\includegraphics[width=1.02\linewidth]{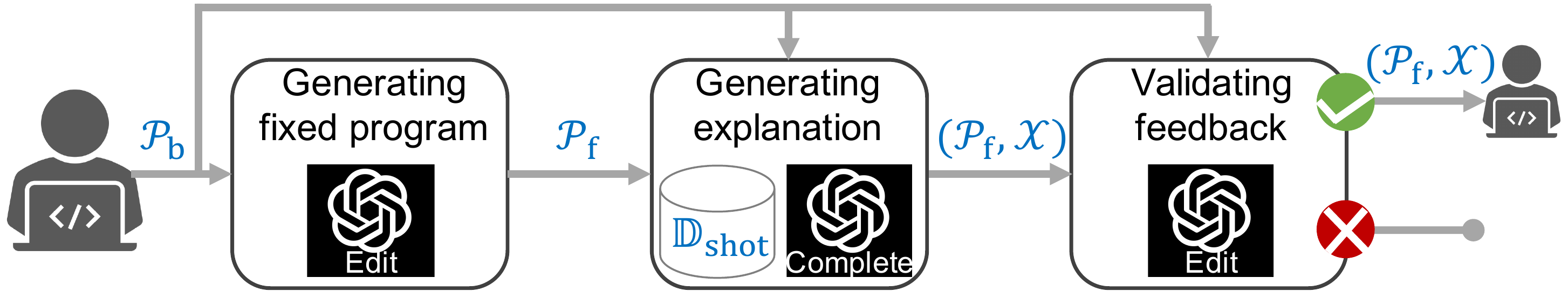}
    \caption{Illustration of three different compoments/stages in \AlgoOurs{}'s feedback generation process; see Section~\ref{sec.methods}.}
	\vspace{-1.5mm}	
	\label{fig.method.pipeline}
\end{figure}

\begin{figure*}[t!]
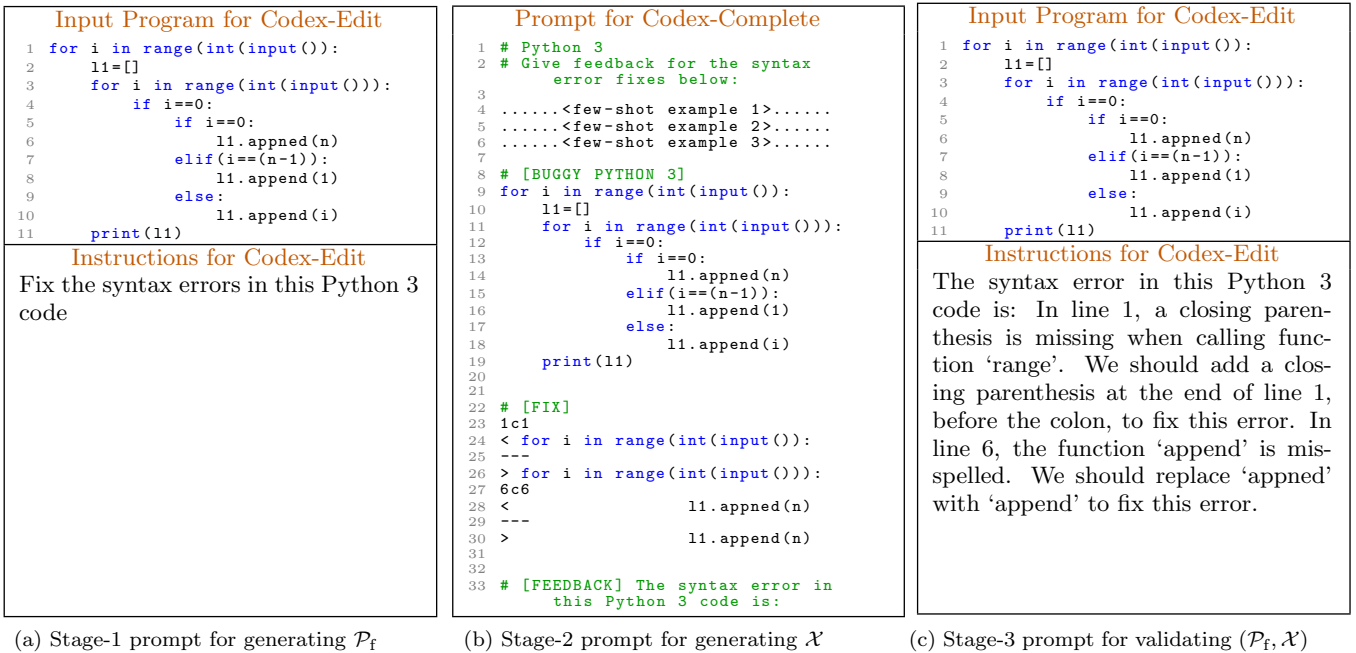

\centering
	\begin{subfigure}[b]{.30\linewidth}
	\centering
	{
\scalebox{1}{
\begin{tabular}{|p{0.001\linewidth}p{0.995\linewidth}p{0.001\linewidth}|}
    \hline
    \multicolumn{3}{|p{1.0\linewidth}|}{
        \centering 
        \textcolor{PromtHeader}{Input Program for Codex-Edit}
    }\\
    &\multicolumn{1}{c}{
\begin{lstlisting}[aboveskip=-0.5em, belowskip=-1.5em, frame=none, basicstyle=\fontsize{6.2}{7}\ttfamily]
for i in range(int(input()):
    l1=[]
    for i in range(int(input())):
        if i==0:
            if i==0:
                l1.appned(n)
            elif(i==(n-1)):
                l1.append(1)
            else:
                l1.append(i)
    print(l1)
\end{lstlisting}
    }&\\
    \hline
    \multicolumn{3}{|p{1.0\linewidth}|}{
        \centering 
        \textcolor{PromtHeader}{Instructions for Codex-Edit}
    }
    \\            
    \multicolumn{3}{|p{1.0\linewidth}|}{
        Fix the syntax errors in this Python 3 code
        \vspace{38.25mm}
    }\\
    \hline
\end{tabular}
}

        \vspace{-0.5mm}
        \caption{Stage-1 prompt for generating \fixedprog{}}
		\label{fig.method.promts.stage1}
	}
	\end{subfigure} 
	\ \ \ \ \ 
	\begin{subfigure}[b]{.30\linewidth}
	\centering
	{
\scalebox{1}{
\begin{tabular}{|p{0.001\linewidth}p{0.995\linewidth}p{0.001\linewidth}|}
    \hline
    \multicolumn{3}{|p{1.0\linewidth}|}{
        \centering 
        \textcolor{PromtHeader}{Prompt for Codex-Complete}
    }\\
    &\multicolumn{1}{p{0.91\linewidth}}{
\begin{lstlisting}[aboveskip=-0.5em, belowskip=-1.5em, frame=none, basicstyle=\fontsize{6.2}{4}\ttfamily]
# Python 3
# Give feedback for the syntax error fixes below:

......<few-shot example 1>......
......<few-shot example 2>......
......<few-shot example 3>......

# [BUGGY PYTHON 3]
for i in range(int(input()):
    l1=[]
    for i in range(int(input())):
        if i==0:
            if i==0:
                l1.appned(n)
            elif(i==(n-1)):
                l1.append(1)
            else:
                l1.append(i)
    print(l1)


# [FIX]
1c1
< for i in range(int(input()):
---
> for i in range(int(input())):
6c6
<                 l1.appned(n)
---
>                 l1.append(n)


# [FEEDBACK] The syntax error in this Python 3 code is:
\end{lstlisting}
    }&\\
    \hline
\end{tabular}
}

        \vspace{-0.5mm}
        \caption{Stage-2 prompt for generating \explaination{}}       
		\label{fig.method.promts.stage2}
	}
	\end{subfigure} 
	\ \ \ \ \ \ \ 
	\begin{subfigure}[b]{.30\linewidth}
	\centering
	{
\scalebox{1}{
\begin{tabular}{|p{0.001\linewidth}p{0.995\linewidth}p{0.001\linewidth}|}
    \hline
    \multicolumn{3}{|p{1.0\linewidth}|}{
        \centering 
        \textcolor{PromtHeader}{Input Program for Codex-Edit}
    }\\
    &\multicolumn{1}{c}{
\begin{lstlisting}[aboveskip=-0.5em, belowskip=-1.5em, frame=none, basicstyle=\fontsize{6.2}{7}\ttfamily]
for i in range(int(input()):
    l1=[]
    for i in range(int(input())):
        if i==0:
            if i==0:
                l1.appned(n)
            elif(i==(n-1)):
                l1.append(1)
            else:
                l1.append(i)
    print(l1)
\end{lstlisting}
    }&\\
    \hline
    \multicolumn{3}{|p{1.0\linewidth}|}{
        \centering 
        \textcolor{PromtHeader}{Instructions for Codex-Edit}
    }
    \\            
    \multicolumn{3}{|p{1.0\linewidth}|}{
        The syntax error in this Python 3 code is: In line 1, a closing parenthesis is missing when calling function `range'. We should add a closing parenthesis at the end of line 1, before the colon, to fix this error. In line 6, the function `append' is misspelled. We should replace `appned' with `append' to fix this error.
        \vspace{12.5mm}
    }\\
    \hline
\end{tabular}
}

        \vspace{-0.5mm}
        \caption{Stage-3 prompt for validating \feedback{}}				
		\label{fig.method.promts.stage3}
	}
	\end{subfigure} 
    \caption{Illustration of prompts used by different stages of \AlgoOurs{} for buggy Python $3$ program in Figure~\ref{fig.intro_overview.cf_pos}. In particular, the ``Instructions for Codex-Edit'' in \textbf{(c)} is obtained by concatenating line33 of \textbf{(b)} and the generated \explaination{} shown in Figure~\ref{fig.intro_overview.cf_pos.exp}.  
    }
	\label{fig.method.promts}
\end{figure*}

\section{Our Technique \AlgoOursTitle{}}\label{sec.methods}
In this section, we present \AlgoOurs{}, our technique to generate high-precision feedback using LLMs. \AlgoOurs{} uses OpenAPI's Codex as LLMs~\cite{DBLP:journals/corr/abs-2107-03374} -- Codex has shown competitive performance on a variety of programming benchmarks~\cite{DBLP:journals/corr/abs-2107-03374,DBLP:conf/ace/Finnie-AnsleyDB22,DBLP:journals/corr/abs-2209-14876,joshi23aaai}, and is particularly suitable for \AlgoOurs{} as we seek to generate both fixed programs and natural language explanations. More specifically, \AlgoOurs{} uses two access points of Codex provided by OpenAI through public APIs: Codex-Edit~\cite{codexedit} and Codex-Complete~\cite{codexcomplete}. As illustrated in Figure~\ref{fig.method.pipeline}, \AlgoOurs{} has the following three components/stages: (1) generating a fixed program \fixedprog{} by editing \buggyprog{} using Codex-Edit; (2) generating natural language explanation \explaination{} using Codex-Complete; (3) validating feedback \feedback{} using Codex-Edit to decide whether the generated feedback is suitable for sharing. The overall pipeline of \AlgoOurs{} is modular and we will evaluate the utility of different components in Section~\ref{sec.experiments}. Next, we provide details for each of these stages.
%

\subsection{Stage-1: Generating Fixed Program}\label{sec.methods.stage1}
Given a student's buggy program \buggyprog{} as input, \AlgoOurs{}'s Stage-1 generates a fixed program \fixedprog{}. We use Codex-Edit for fixing/repairing the buggy program in this stage since it has shown to be competitive in program repair benchmarks in recent works~\cite{fan2022improving}. Figure~\ref{fig.method.promts.stage1} shows a sample prompt used by \AlgoOurs{} to query Codex-Edit for the buggy Python $3$ program in Figure~\ref{fig.intro_overview.cf_pos.buggy}. The process of generating \fixedprog{} is determined by two hyperparameters: (i) $t_1 \in [0.0, 1.0]$ is the temperature value specified when querying Codex-Edit and controls stochasticity/diversity in generated programs; (ii) $n_1$ controls the number of queries made to Codex-Edit.

More concretely, \AlgoOurs{} begins by making $n_1$ queries to Codex-Edit with temperature $t_1$. Then, out of $n_1$ generated programs, \AlgoOurs{} selects \fixedprog{} as the program that is syntactically correct and has the smallest \emph{edit-distance} to \buggyprog{}. Here, edit-distance between two programs is measured by first tokenizing programs using Pygments library~\cite{pygments} and then computing Levenshtein edit-distance over token strings.\footnote{Note that buggy programs are not parseable to \emph{Abstract Syntax Tree} (AST) representations and string-based distance is commonly used in such settings (e.g., see~\cite{DBLP:journals/corr/abs-2209-14876}).\label{footnote.codedistance}}
If Stage-1 is unable to generate a fixed program, the process stops without generating any feedback; see Footnote~\ref{footnote.nofeedback}. In our experiments, we set $(t_1=0.5, n_1=10)$ and obtained a high success rate of generating a fixed program \fixedprog{} with a small number of edits w.r.t. \buggyprog{}.
%

\begin{figure*}[t!]
\centering
    \begin{subfigure}[b]{.48\linewidth}
	\centering
	{
\scalebox{0.82}{
\setlength\tabcolsep{3.2pt}
\renewcommand{\arraystretch}{1.0}
\begin{tabular}{l||p{5cm}p{5cm}||p{5cm}p{5cm}}
    \toprule
    \textbf{Technique} & \multicolumn{2}{c||}{\textbf{TigerJython}} & \multicolumn{2}{c}{\textbf{Codeforces}} \\
    &   \multicolumn{1}{c}{Precision} &  \multicolumn{1}{c||}{Coverage} &   \multicolumn{1}{c}{Precision} &  \multicolumn{1}{c}{Coverage} \\      
    \midrule
    \AlgoPEM  & \multicolumn{1}{l}{$05.0~(1.0)$} & \multicolumn{1}{l||}{$92.5~(1.6)$} & \multicolumn{1}{l}{$35.0~(2.4)$} & \multicolumn{1}{l}{$98.8~(0.8)$} \\
    \midrule
    \AlgoPyFiXNone  & \multicolumn{1}{l}{$00.9~(0.5)$} & \multicolumn{1}{l||}{$92.5~(1.6)$} & \multicolumn{1}{l}{$03.0~(0.4)$} & \multicolumn{1}{l}{$98.8~(0.8)$} \\
    \AlgoPyFiXRand  & \multicolumn{1}{l}{$21.6~(1.7)$} & \multicolumn{1}{l||}{$92.5~(1.6)$} & \multicolumn{1}{l}{$48.5~(2.6)$} & \multicolumn{1}{l}{$98.8~(0.8)$} \\
    \AlgoPyFiXSel  & \multicolumn{1}{l}{$38.9~(3.5)$} & \multicolumn{1}{l||}{$92.5~(1.6)$} & \multicolumn{1}{l}{$55.2~(3.9)$} & \multicolumn{1}{l}{$98.8~(0.8)$} \\
    \midrule
    \AlgoPyFiXParallelSel  & \multicolumn{1}{l}{$15.8~(1.8)$} & \multicolumn{1}{l||}{$92.5~(1.6)$} & \multicolumn{1}{l}{$15.6~(2.8)$} & \multicolumn{1}{l}{$98.8~(0.8)$} \\
    \midrule
    %
    %
    $\AlgoPyFiXVRule_{\precision{}\geq70}$  & \multicolumn{1}{l}{$48.6~(4.4)$} & \multicolumn{1}{l||}{$30.8~(12.5)$} & \multicolumn{1}{l}{$61.6~(9.0)$} & \multicolumn{1}{l}{$38.3~(10.5)$} \\
    \cellcolor{green!15}$\AlgoOurs_{\precision{}\geq70}$  & \multicolumn{1}{l}{\cellcolor{green!15}$76.0~(4.0)$} & \multicolumn{1}{l||}{\cellcolor{green!15}$31.2~(4.0)$} & \multicolumn{1}{l}{\cellcolor{green!15}$72.4~(6.2)$} & \multicolumn{1}{l}{\cellcolor{green!15}$64.2~(6.3)$} \\
    $\AlgoPyFiXVOracle_{\precision{}\approx{}V_{\precision{}\geq70}}$  & \multicolumn{1}{l}{$76.1~(0.4)$} & \multicolumn{1}{l||}{$47.1~(3.4)$} & \multicolumn{1}{l}{$72.8~(0.1)$} & \multicolumn{1}{l}{$75.0~(5.7)$} \\
    \bottomrule   
\end{tabular}
}
		\caption{Results for different techniques, reported as mean (stderr)}			
		\label{fig.experiments.table}
	}
	\end{subfigure}
    \ \  
	\begin{subfigure}[b]{.24\linewidth}
	\centering
	{
		\includegraphics[width=0.91\linewidth]{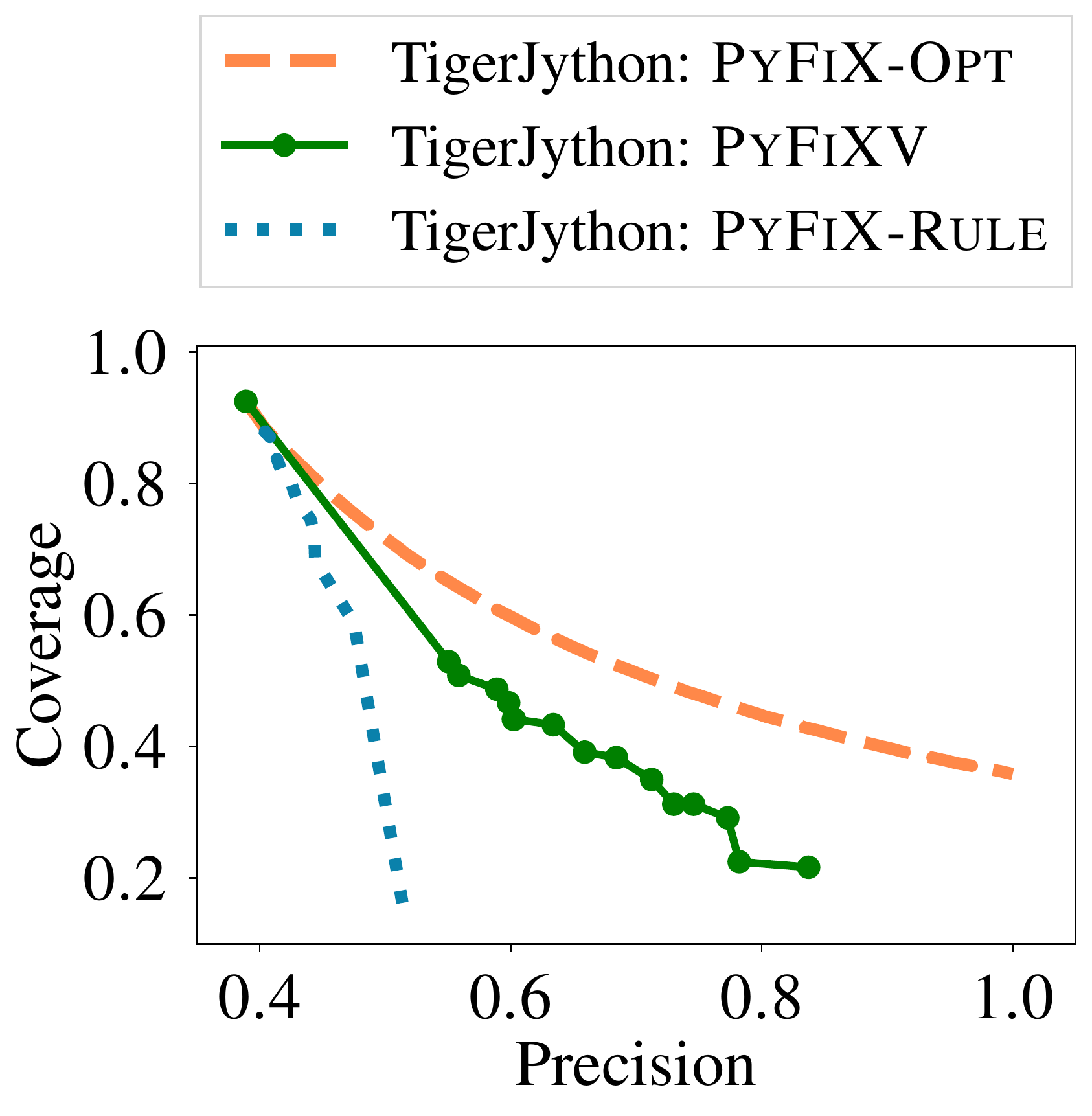}
        \vspace{-1.8mm}
		\caption{TigerJython trade-off curve}				
		\label{fig.experiments.precision.tj}
	}
	\end{subfigure}
    \ \ 
	\begin{subfigure}[b]{.24\linewidth}
	\centering
	{
		\includegraphics[width=0.91\linewidth]{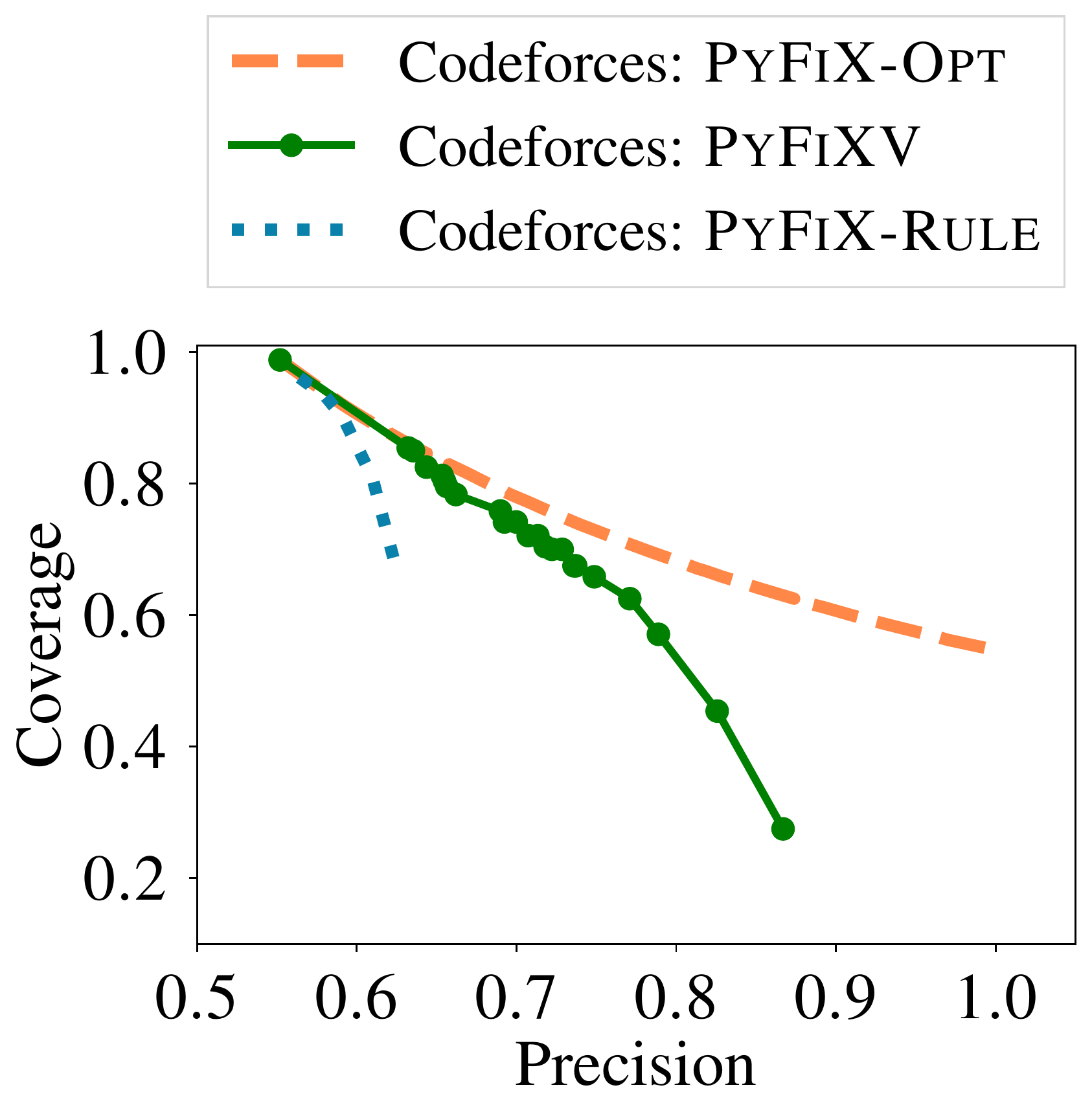}
        \vspace{-1.8mm}
        \caption{Codeforces trade-off curve}				
		\label{fig.experiments.precision.cf}
	}
	\end{subfigure}  
    \caption{Experimental results on two real-world datasets of Python programs, namely TigerJython~\cite{DBLP:conf/sigcse/KohnM20} and Codeforces~\cite{codeforces}.}
	\label{fig.experiments}
\end{figure*}

\subsection{Stage-2: Generating Explanation}\label{sec.methods.stage2}
Given \buggyprog{} and \fixedprog{} as inputs, \AlgoOurs{}'s Stage-2 generates a natural language explanation \explaination{} describing errors/fixes. We use Codex-Complete in this stage as it is naturally suited to generate text by completing a prompt~\cite{DBLP:journals/corr/abs-2107-03374,macneil23sigcse,leinonen23sigcse}. 
A crucial ingredient of Stage-2 is the annotated dataset \dataFewS{} used to select few-shot examples when querying Codex-Complete (see Figure~\ref{fig.method.pipeline}).
Figure~\ref{fig.method.promts.stage2} shows a sample prompt used by \AlgoOurs{} to query Codex-Complete for the scenario in Figure~\ref{fig.intro_overview.cf_pos}. In Figure~\ref{fig.method.promts.stage2}, line4--line6 indicate three few-shot examples (not shown for conciseness), line9--line19 provides \buggyprog{}, line23--line30 provides \fixedprog{} in the form of line-diff w.r.t. \buggyprog{}, and line33 is the instruction to be completed by Codex-Complete. Given a prompt, the process of generating \explaination{} is determined by two hyperparameters: (i) a temperature value $t_2$ ($=0$) and (ii) the number of queries $n_2$ ($=1$). Next, we discuss the role of \dataFewS{} in selecting few-shots examples.

%
When querying Codex-Complete, we use three few-shot examples selected from \dataFewS{}, an annotated dataset of examples comprising buggy programs and desired feedback obtained by expert annotations (see Section~\ref{sec.experiments.data}). These annotated examples essentially provide a context to LLMs and have shown to play an important role in optimizing the generated output (e.g., see \cite{DBLP:journals/corr/abs-2107-03374,DBLP:conf/nips/BrownMRSKDNSSAA20,DBLP:journals/pacmpl/BavishiJCFGLRT22,DBLP:journals/corr/abs-2209-14876,joshi23aaai}). In our case, \dataFewS{} provides contextualized training data, capturing the format of how experts/tutors give explanations. Given \buggyprog{} and \fixedprog{}, we use two main criteria to select few-shot examples. The primary criterion is to pick examples where the error type of buggy program in the example is same as that of \buggyprog{}---the underlying parser/compiler provides error types (e.g., `InvalidSyntax', `UnexpectedIndent'). The secondary criterion (used to break ties in the selection process) is based on the edit-distance \emph{between} the \emph{diff} of buggy/fixed program in the example and \emph{diff} of \buggyprog{}/\fixedprog{}. In Section~\ref{sec.experiments}, we conduct ablations to showcase the importance of selecting few-shots.

\subsection{Stage-3: Validating Feedback}\label{sec.methods.stage3}
Given \buggyprog{} and \feedback{} as inputs, \AlgoOurs{}'s Stage-3 validates the feedback quality and makes a binary decision of ``accept'' (feedback is suitable for sharing) or ``reject'' (feedback is discarded). \AlgoOurs{} uses a novel run-time feedback validation mechanism using Codex-Edit to decide whether the feedback \feedback{} w.r.t. \buggyprog{} is of good quality. Here, Codex-Edit is used in the flipped role of a \emph{simulated student model} -- the intuition is that a good quality explanation \explaination{}, when provided in Codex-Edit's prompt instruction, should increase Codex-Edit's success in converting \buggyprog{} to \fixedprog{}.
Figure~\ref{fig.method.promts.stage3} shows a sample prompt used by \AlgoOurs{} to query Codex-Edit for the scenario in Figure~\ref{fig.intro_overview.cf_pos}---see the caption on how ``Instructions for Codex-Edit'' in Figure~\ref{fig.method.promts.stage3} is obtained.\footnote{In our initial experiments, we tried using alternative signals for validation, such as (a) Codex-Complete's probabilities associated with generated \explaination{}; (b) automatic scoring of \explaination{} w.r.t. explanations in few-shots using BLEU score~\cite{DBLP:conf/acl/PapineniRWZ02}; (c) filtering based on \explaination{}'s length. Section~\ref{sec.experiments} reports results for (c)  as it had the highest performance among these alternatives.\label{footnote.alternatevalidation}}

The validation mechanism has three hyperparameters: (i) $t_3 \in [0.0, 1.0]$ is the temperature value specified when querying Codex-Edit; (ii) $n_3$ controls the number of queries made to Codex-Edit; (iii) $h_3 \in [1, n_3]$ is the threshold used for acceptance decision. More concretely, \AlgoOurs{} begins by making $n_3$ queries to Codex-Edit with temperature $t_3$. Then, out of $n_3$ generated programs, \AlgoOurs{} counts the number of programs that don't have syntax errors and have an \emph{exact-match} with \fixedprog{}. Here, exact-match is checked by converting programs to their \emph{Abstract Syntax Tree} (AST)-based normalized representations.\footnote{We check for AST-based exact match instead of checking for Levenshtein edit-distance over token strings being $0$ (see Section~\ref{sec.methods.stage1}). AST-based exact match is more relaxed than edit-distance being $0$ -- AST-based representation ignores certain differences between codes, e.g., based on extra spaces and comments. We used the AST-based exact match in the validation mechanism as it is more robust to such differences.}
Finally, the validation mechanism accepts the feedback if the number of exact matches is at least $h_3$. These hyperparameters $(t_3, n_3, h_3)$ also provide a precision knob and are selected to obtain the desired precision level, as discussed next.



\subsection{Precision and Coverage Trade-Off}\label{sec.methods.knob}
\AlgoOurs{}'s validation mechanism provides a precision knob to control the precision and coverage trade-off (see 
 performance metrics in Section~\ref{sec.problem.objectives}). Let \precision{} be the desired precision level we want to achieve for \AlgoOurs{}. The idea is to choose Stage-3 hyperparameters $(t_3, n_3, h_3)$ that achieve \precision{} precision level. For this purpose, we use a calibration dataset \dataVal{} for picking the hyperparameters. More concretely, in our experiments, \AlgoOurs{} first computes performance metrics on \dataVal{} for the following range of  values: (i) $t_3 \in \{0.3, 0.5, 0.8\}$; (ii) $n_3 \in \{10\}$; (iii) $h_3 \in \{1, 2, \ldots, 10\}$. Then, it chooses $(t_3, n_3, h_3)$ that has at least \precision{} precision level and maximizes coverage; when achieving the desired \precision{} is not possible, then the next lower possible precision is considered. The chosen values of hyperparameters are then used in \AlgoOurs{}'s Stage-3 validation mechanism. We refer to $\AlgoOurs_{\precision{}\geq x}$ as the version of \AlgoOurs{} calibrated with $\precision{} \geq x$.


\begin{figure*}[t!]
\centering
	\begin{subfigure}[b]{0.34\linewidth}
	\centering
	{
\begin{lstlisting}[basicstyle=\fontsize{6.5}{5.8}\ttfamily]
import time

def cd(seconds):
    while seconds > 0:
        rem_m = seconds // 60
        rem_s = seconds % 60
        remaining_time = str(rem_m) + " minutes " + str(rem_s) + " seconds"
        print(remaining_time, end='\r')
        seconds = seconds - 1
        time.sleep(1)
    
    print("Xxxxx xxxxX")

cd(60)
\end{lstlisting}
        \vspace{-2mm}
		\caption{Student's buggy program \buggyprog{}}				
		\label{fig.intro_overview.tj_poscomplex.buggy}
	}
	\end{subfigure}
	\quad \ \ \ 
	\begin{subfigure}[b]{.34\linewidth}
	\centering
	{
\begin{lstlisting}[basicstyle=\fontsize{6.5}{5.0}\ttfamily,
        linebackgroundcolor={
            \ifnum\value{lstnumber}=1\color{CodeHighlight}\fi
        }]
from __future__ import print_function
import time

def cd(seconds):
    while seconds > 0:
        rem_m = seconds // 60
        rem_s = seconds % 60
        remaining_time = str(rem_m) + " minutes " + str(rem_s) + " seconds"
        print(remaining_time, end='\r')
        seconds = seconds - 1
        time.sleep(1)
    
    print("Xxxxx xxxxX")

cd(60)
\end{lstlisting}
        \vspace{-2mm}
		\caption{Generated \fixedprog{}}				
		\label{fig.intro_overview.tj_poscomplex.fixed}
	}
	\end{subfigure}	
	\ 
	\begin{subfigure}[b]{.14\linewidth}
	\centering
	{
         \scalebox{0.83}{
        \begin{tabular}{|p{1\linewidth}|}
     	  \hline
            \multicolumn{1}{|p{1\linewidth}|}{
                  The student wants to use the print function from Python 3. To do so, the correct way is to add the line `from \_\_future\_\_ import print\_function' at the beginning of the code. 
                  \vspace{0.75mm}
            }\\
            \hline
        \end{tabular}
        }
        \caption{Generated \explaination{}}				
		\label{fig.intro_overview.tj_poscomplex.exp}
	}
	\end{subfigure}
    \ \
	\begin{subfigure}[b]{.105\linewidth}
	\centering
	{
		\includegraphics[height=3.5cm]{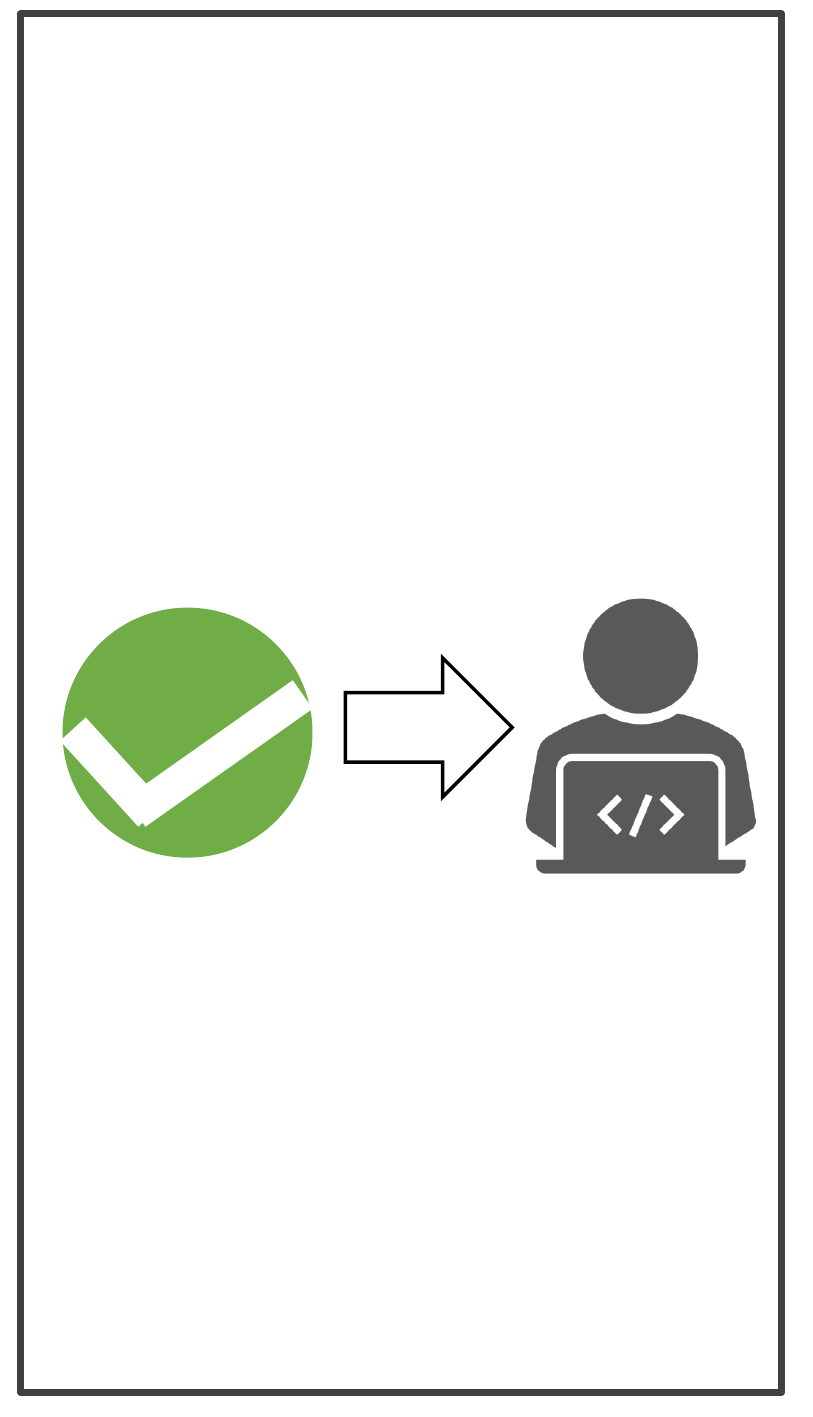}
        \vspace{-4.1mm}
		\caption{Validation}				
		\label{fig.intro_overview.tj_poscomplex.validation}
	}
	\end{subfigure}
	\vspace{-2mm}   
    \caption{Similar to Figure~\ref{fig.intro_overview.tj_pos}, this illustrative example showcases \AlgoOurs{} on a buggy Python $2$ program from TigerJython~\cite{DBLP:conf/sigcse/KohnM20}. 
    }
	\label{fig.intro_overview.tj_poscomplex}
\end{figure*}

\begin{figure*}[t!]
\centering
	\begin{subfigure}[b]{.30\linewidth}
	\centering
	{
\begin{lstlisting}[basicstyle=\fontsize{6.5}{11.9}\ttfamily]
name = input("Xxx xx?")
num = input("Xxx xx xxxxxxx?")
print ("Xxxx " + name + "Xxx ")*,num
\end{lstlisting}
        \vspace{-2mm}
		\caption{Student's buggy program \buggyprog{}}				
		\label{fig.intro_overview.tj_neg.buggy}
	}
	\end{subfigure}
	\quad \ \ \  
	\begin{subfigure}[b]{.295\linewidth}
	\centering
	{
\begin{lstlisting}[basicstyle=\fontsize{6.5}{11.9}\ttfamily,
        linebackgroundcolor={
            \ifnum\value{lstnumber}=3\color{CodeHighlight}\fi
        }]
name = input("Xxx xx?")
num = input("Xxx xx xxxxxxx?")
print ("Xxxx " + name + "Xxx ")*num
\end{lstlisting}
        \vspace{-2mm}        
		\caption{Generated \fixedprog{}}				
		\label{fig.intro_overview.tj_neg.fixed}
	}
	\end{subfigure}
    \  
	\begin{subfigure}[b]{.236\linewidth}
	\centering
	{
         \scalebox{0.83}{
        \begin{tabular}{|p{1\linewidth}|}
     	  \hline
            \multicolumn{1}{|p{1\linewidth}|}{
                \textcolor{ExpHighlight}{The student forgets to enclose a string literal with quotes. We can fix the error by enclosing the string literal in line 3 with a pair of double quotes.}
                \vspace{-4mm}
            }\\\\
            \hline
        \end{tabular}
        }
        \caption{Generated \explaination{}}				
		\label{fig.intro_overview.tj_neg.exp}
	}
	\end{subfigure}
	\begin{subfigure}[b]{.11\linewidth}
	\centering
	{
        \includegraphics[height=1.58cm]{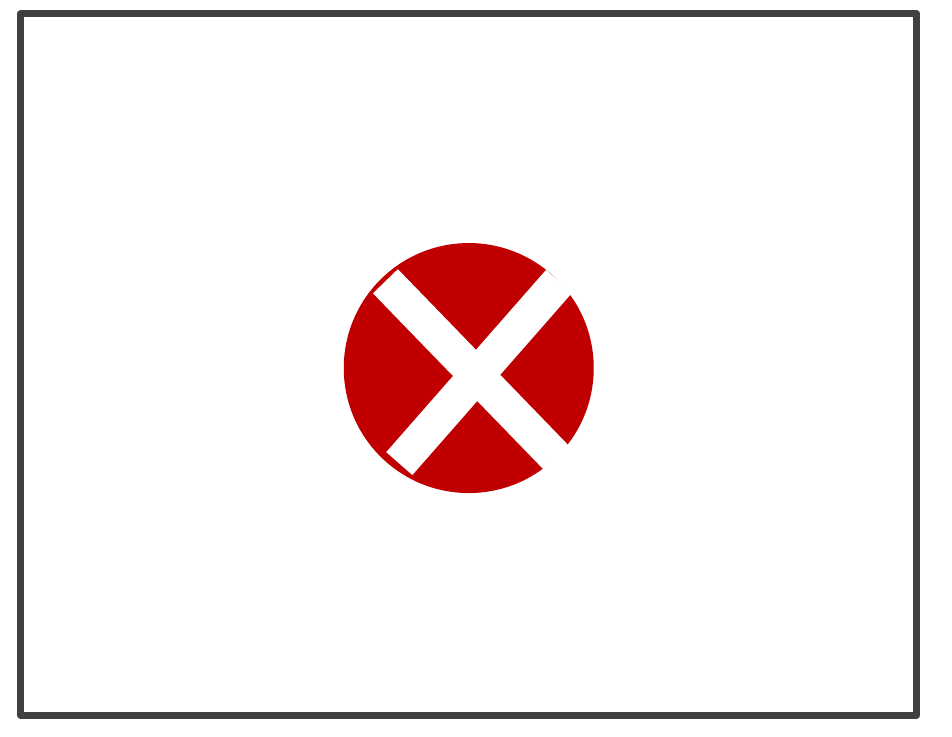}
        \vspace{-4.0mm}
		\caption{Validation}				
		 \label{fig.intro_overview.tj_neg.validation}
	}
	\end{subfigure}
    \caption{Similar to Figure~\ref{fig.intro_overview.cf_neg}, this example showcases \AlgoOurs{} on a buggy Python $2$ program from TigerJython~\cite{DBLP:conf/sigcse/KohnM20}. 
    \AlgoOurs{}'s validation mechanism \emph{successfully rejected} the generated feedback (we marked text in \textbf{(c)} to highlight issues with explanation).
    }
	\label{fig.intro_overview.tj_neg}
\end{figure*}

\section{Experimental Evaluation}\label{sec.experiments}
We perform evaluations using two real-world Python programming datasets, namely TigerJython~\cite{DBLP:conf/sigcse/KohnM20} and Codeforces \cite{codeforces}. We picked Python because of its growing popularity as an introductory programming language; notably, \AlgoOurs{} can be used with other languages by appropriately changing the prompts and tokenizers used. We use OpenAI's public APIs for Codex-Edit~\cite{codexedit} (\emph{model=code-davinci-edit-001}) and Codex-Complete~\cite{codexcomplete} (\emph{model=code-davinci-002}). We begin by describing 
 different techniques used in the evaluation.
%

\subsection{Baselines and Variants of \AlgoOursTitle{}}
\label{sec.experiments.baselines}

\textbf{Default programming-error-messages without validation.} As our first baseline, \AlgoPEM{} uses \AlgoOurs{}'s Stage-1 to generate \fixedprog{} and uses programming-error-messages provided by the programming environment as  \explaination{}. \AlgoPEM{} uses error messages provided by Python $2.7$ environment for TigerJython and Python $3.12$ environment for Codeforces. This baseline is without validation (i.e., the generated feedback is always accepted).%

\textbf{Variants of \AlgoOurs{} without validation.} \AlgoPyFiXSel{} is a variant of \AlgoOurs{} without the validation mechanism (i.e., only uses Stage-1 and Stage-2). \AlgoPyFiXRand{} is a variant of \AlgoPyFiXSel{} where few-shot examples in Stage-2 are picked randomly from \dataFewS{}. \AlgoPyFiXNone{} is a variant of \AlgoPyFiXSel{} that doesn't use few-shot examples in Stage-2. \AlgoPyFiXParallelSel{} is a variant of \AlgoPyFiXSel{} that runs Stage-1 and Stage-2 in parallel; hence, Stage-2's prompt doesn't make use of \fixedprog{}. All these variants are without validation (i.e., the generated feedback is always accepted).

\textbf{Techniques with alternative validation mechanisms.} We consider two variants of \AlgoOurs{}, namely \AlgoPyFiXVRule{} and \AlgoPyFiXVOracle{}, that use different validation mechanisms (i.e., replace \AlgoOurs{}'s Stage-3 with an alternative validation). \AlgoPyFiXVRule{} validates \feedback{} based on \explaination{}'s length, as noted in Footnote~\ref{footnote.alternatevalidation}. Given a  hyperparameter $h_{r}$, \feedback{} is accepted if the number of tokens in \explaination{} is at most $h_{r}$, where tokenization is done by splitting on whitespaces/punctuations. \AlgoPyFiXVRule{}'s  $h_{r}$ is picked from the set $\{30, 40, 50, \ldots, 200\}$ based on the desired precision level \precision{}, by following the calibration process in Section~\ref{sec.methods.knob}. \AlgoPyFiXVOracle{} uses an oracle validation that has access to expert's ratings for the generated feedback \feedback{}. Then, for a desired \precision{}, \AlgoPyFiXVOracle{} performs optimal validation and highlights the maximum coverage achievable on \dataTest{} for the generated feedback.

%


\subsection{Datasets and Evaluation Procedure} \label{sec.experiments.data}

\textbf{Datasets and annotations for few-shot examples.} As our first dataset, namely TigerJython, we have $240$ distinct  Python $2$ programs written by students in TigerJython's educational programming environment~\cite{DBLP:conf/sigcse/KohnM20}. We obtained a private and anonymized version of the dataset used in~\cite{DBLP:conf/sigcse/Kohn19}, with string literals in programs replaced with sequences of `x' (e.g., see Figure~\ref{fig.intro_overview.tj_pos}). As our second dataset, namely Codeforces, we curated $240$ distinct Python $3$ programs from the Codeforces website using their public APIs~\cite{codeforces}, inspired by similar works that curate Codeforces dataset~\cite{Caballero_Description2Code_Dataset_2016,li_competition-level}. Programs in both datasets have syntax errors and have token length at most $500$ (see Section~\ref{sec.methods.stage1} about program tokenization). For the Codeforces dataset, we only include programs submitted to contests held from July 2021 onwards (after the cut-off date for Codex's training data~\cite{DBLP:journals/corr/abs-2107-03374}). Since a part of these datasets will be used for few-shot examples (as \dataFewS{} in \AlgoOurs{}'s Stage-2), we asked experts to annotate these $480$ programs with feedback (i.e., a fixed program along with an explanation). Three experts, with extensive experience in Python programming and tutoring, provided annotations.

\textbf{Evaluation procedure and feedback ratings.} 
Given a dataset \data{} with $240$ buggy programs, we can evaluate a technique by splitting \data{} as follows: (a) \dataTest{} ($25$\%) for reporting precision and coverage performance metrics; (b) \dataFewS{} ($50$\%) for few-shot examples; (c) \dataVal{} ($25$\%) for calibrating validation mechanism. To report overall performance for techniques, we perform a cross-validation procedure with four evaluation rounds while ensuring that \dataTest{} across four rounds are non-overlapping. We then report aggregated results across these rounds as average mean (stderr). As discussed in Sections~\ref{sec.problem.preliminaries}~and~\ref{sec.problem.objectives}, evaluating these performance metrics requires feedback ratings by experts to assess the quality of the feedback generated by each technique.\footnote{We note that precision and coverage performance metrics for different techniques are reported for the end-to-end process associated with each technique, and not just for the validation mechanism. Also, even if a technique doesn't use any validation mechanism, the coverage could be less than $100.0$ as discussed in Footnote~\ref{footnote.nofeedback}.} For example, evaluating metrics on TigerJython dataset for \AlgoOurs{}  requires $480$ feedback ratings ($4 \times 60$ for \dataTest{} and $4 \times 60$ for \dataVal{}). 
To begin, we did a smaller scale investigation to establish the rating criteria, where two experts rated $100$ generated feedback instances; we obtained Cohen's kappa reliability value $0.72$ indicating \emph{substantial agreement} between experts~\cite{warrens2015five}. Afterward, one expert (with experience in tutoring Python programming classes) did these feedback ratings for the evaluation results.\footnote{We note that the experts were blinded to the condition (technique) associated with each feedback instance when providing ratings. Moreover, these generated feedback instances were given to experts in randomized order across conditions instead of grouping them per condition.}
%

\subsection{Results}
\textbf{Comparison of different techniques.} Figure~\ref{fig.experiments.table} provides a comparison of different techniques on two datasets. All techniques here use \AlgoOurs{}'s Stage-1 to obtain \fixedprog{}. The coverage numbers of $92.5$ and $98.8$ reported in Figure~\ref{fig.experiments.table} correspond to the success rate of obtaining \fixedprog{} on these datasets (the average edit-distance between \buggyprog{} and \fixedprog{} is about $10.4$ and $7.5$ tokens on these datasets, respectively). For our baseline \AlgoPEM{}, we see a big jump in precision from $5.0$ for TigerJython (Python $2$) to $35.0$ for Codeforces (Python $3$), owing to enhanced error messages in recent Python versions~\cite{python3.10,python3.11,python3.12}. Results for $\AlgoOurs_{\precision{}\geq70}$ in comparison with results for \AlgoPyFiXSel{}, \AlgoPyFiXRand{}, \AlgoPyFiXNone{}, and \AlgoPyFiXParallelSel{} showcase the utility of different components used in \AlgoOurs{}'s pipeline. Comparing $\AlgoOurs_{\precision{}\geq70}$ with $\AlgoPyFiXVRule_{\precision{}\geq70}$ shows that \AlgoOurs{}'s validation substantially outperforms \AlgoPyFiXVRule{}'s validation.\footnote{
When comparing $\AlgoOurs_{\precision{}\geq70}$ with these techniques in Figure~\ref{fig.experiments.table}, the results are significantly different w.r.t. $\chi^2$ tests~\cite{cochran1952chi2} ($p\leq0.0001$); here, we use contingency tables with two rows (techniques) and four columns ($240$ data points mapped to four possible precision/coverage outcomes).} Lastly, results for $\AlgoPyFiXVOracle_{\precision{}\approx{}V_{\precision{}\geq70}}$ are obtained by setting the desired precision level for \AlgoPyFiXVOracle{} to match that of $\AlgoOurs_{\precision{}\geq70}$ on \dataTest{} -- the coverage numbers ($47.1$ for TigerJython and $75.0$ for Codeforces) indicate the maximum possible achievable coverage. Notably, $\AlgoOurs_{\precision{}\geq70}$ achieves a competitive coverage of $64.2$ on Codeforces.\footnote{\looseness-1Techniques \AlgoPyFiXSel{}, \AlgoPyFiXVRule{}, $\AlgoOurs_{\precision{}\geq70}$, and  $\AlgoPyFiXVOracle_{\precision{}\approx{}V_{\precision{}\geq70}}$ differ only in terms of validation mechanisms. We can compare the validation mechanisms used in these techniques based on F1-score. The F1-scores of these four techniques are as follows: $0.56$, $0.39$, $0.70$, and $0.86$ for TigerJython, respectively; $0.71$, $0.47$, $0.77$, and $0.84$ for Codeforces, respectively.}
%
%



\textbf{Precision and coverage trade-off curves.} %
The curves in Figures~\ref{fig.experiments.precision.tj}~and~\ref{fig.experiments.precision.cf} are obtained by picking different desired precision levels \precision{} and then computing precision/coverage values on \dataTest{} w.r.t. \precision{}.
The curves for \AlgoPyFiXVOracle{} show the maximum possible coverage achievable on \dataTest{} for different precision levels \precision{} using our generated feedback. 
To obtain these curves for \AlgoOurs{} and \AlgoPyFiXVRule{}, we did calibration directly on \dataTest{} instead of \dataVal{} (i.e., doing ideal calibration for their validation mechanisms when comparing with \AlgoPyFiXVOracle{}'s curves). These curves highlight the precision and coverage trade-off offered by \AlgoOurs{} in comparison to a simple rule-based validation and the oracle validation.
%
%

\textbf{Qualitative analysis.} We have provided several illustrative examples to demonstrate our technique \AlgoOurs{}. Figures~\ref{fig.intro_overview.tj_pos}, \ref{fig.intro_overview.cf_pos},~and~\ref{fig.intro_overview.tj_poscomplex} show examples where \AlgoOurs{}'s Stage-1 and Stage-2 generate good quality feedback and Stage-3 successfully accepts the feedback. Figures~\ref{fig.intro_overview.cf_neg}~and~\ref{fig.intro_overview.tj_neg} show examples where \AlgoOurs{}'s Stage-1 and Stage-2 generate bad quality feedback and Stage-3 successfully rejects the feedback. Figure~\ref{fig.intro_overview.tj_poscomplex} highlights that \AlgoOurs{} can make non-trivial fixes in the buggy program and correctly explain them in a comprehensible way. 
Figure~\ref{fig.intro_overview.cf_neg} shows an example where the overall feedback is bad quality and successfully rejected, though parts of the generated explanation are correct; this could potentially be useful for tutors in a human-in-the-loop approach.

%

\section{Concluding Discussions}\label{sec.conclusions}
We investigated using LLMs to generate feedback for fixing programming syntax errors. In particular, we considered feedback in the form of a fixed program along with a natural language explanation. We focussed on the challenge of generating high-precision feedback, which is crucial before deploying such technology in classrooms. Our proposed technique, \AlgoOurs, ensures high precision through a novel run-time validation mechanism and also provides a precision knob to educators. We performed an extensive evaluation to showcase the efficacy of \AlgoOurs{} on two real-world Python programming datasets. There are several interesting directions for future work, including (a) improving \AlgoOurs{}'s components to obtain better precision/coverage trade-off, e.g., by adapting our technique to use recent LLMs such as ChatGPT~\cite{ChatGPT} and GPT-4~\cite{GPT4} instead of Codex; (b) extending \AlgoOurs{} beyond syntax errors to provide feedback for programs with semantic errors or partial programs; (c) incorporating additional signals in \AlgoOurs{}'s validation mechanism; (d) conducting real-world studies in classrooms.
%

\section{Acknowledgments}
Funded/Co-funded by the European Union (ERC, TOPS, 101039090). Views and opinions expressed are however those of the author(s) only and do not necessarily reflect those of the European Union or the European Research Council. Neither the European Union nor the granting authority can be held responsible for them.

\bibliographystyle{unsrtnat}
\bibliography{main}


\end{document}